\documentstyle[preprint,epsfig,epsf,aps,floats]{revtex}
\newcommand{\refe}[1]{(\ref{#1})}
\newcommand{\re}{{\mathrm Re}}
\newcommand{\im}{{\mathrm Im}}

\begin{document}  
\newlength{\captsize} \let\captsize=\small 
\newlength{\captwidth}                     

\tightenlines

\title{A study on quantum decoherence phenomena with three 
generations of neutrinos}  

\preprint{\vbox{\hbox{IFUSP-DFN/076-2002}}}

\author{A.\ M.\ Gago$^{1,3}$~\thanks{Email address: agago@charme.if.usp.br},
E.\ M.\ Santos$^{2}$~\thanks{Email address: emoura@charme.if.usp.br}, W.\ J. \ C. Teves$^{2}$~\thanks{Email address: teves@charme.if.usp.br} and 
R.\ Zukanovich Funchal$^{2}$~\thanks{Email address: zukanov@charme.if.usp.br}\\}


\address{\sl 
$^1$ Department of Physics, California State University, Dominguez Hills, Carson CA 90747, USA\\
$^2$ Instituto de F\'\i sica, Universidade de S\~ao Paulo,
    C.\ P.\ 66.318, 05315-970\\  S\~ao Paulo, Brazil\\
$^3$ Secci\'on F\'{\i}sica, Departamento de Ciencias,
    Pontificia Universidad Cat\'{o}lica del Per\'{u}, Apartado 1761, Lima, Per\'{u}\\
\vglue -0.2cm
}    
\maketitle

\vspace{-.2cm}


\begin{abstract}
\noindent
\vglue -1.1cm
 Using the open quantum system approach applied to the neutrino system, 
 we derive three generations neutrino probability formulae considering 
 the oscillation induced by mass plus quantum decoherence contributions. 
 The introduction of these dissipative effects is done through the quantum 
 dynamical semigroup formalism. In addition to the theoretical interest of 
 the approach, at least from the completeness point of view, this extension 
 of the formalism to the three flavors, provide us with a direct application: 
 we can analyze qualitatively the consistency of the two generation pure 
 decoherence solution to the atmospheric neutrino problem, accommodated within 
 this enlarged scheme, with the mean tendencies observed for some of the current neutrino experimental data. This study was performed based on different 
 choices of the $3 \times 3$ mixing matrix selected in order to 
 adjust the $P_{\nu_{\mu} \rightarrow \nu_{\mu}}$ to the same form it has 
 for the decoherence solution in two generations. Our qualitative tests 
 for decoherence with three neutrinos show a clear incompatibility between  
 neutrino data and the theoretical expectations.   
\end{abstract}
\pacs{PACS numbers: }

\section{Introduction}
\label{sec:intro}

 In the open quantum system approach\cite{Davies1}, the evolution of a 
 system interacting with an environment is described in an effective way, 
 that is, the interaction with the environment is incorporated in the 
 description of the evolution of the system. In general, the effects 
 produced by the interaction cause dissipation and irreversibility. 
 This treatment was originally developed for quantum optics\cite{Louisell}, 
 in order to take into account the system-reservoir (environment) interaction, 
 and it has already been applied to elementary particle systems. 
 Recently, it has also been used in the study of two neutrino 
 oscillations\cite{Benatti1,Benatti2}, modifying the well known oscillation 
 probability, due to the presence of dissipative effects. 
 One very interesting feature of the oscillation probability in this situation 
 is that even in the case that neutrinos are massless, we can have neutrino 
 flavor conversion\cite{Benatti2,gstz1,gstz2}. It has been pointed out that 
 this mechanism is able to explain well, in the context of two generations, 
 the atmospheric neutrino data collected by the Super-Kamiokande (SK) 
 experiment\cite{lisi}, as long as the damping parameter is 
 $\propto 1/ \mbox{E}_{\nu}$, $E_\nu$ being the neutrino energy.

 The main goal of this paper is to extend this formalism of mass, mixing and 
 quantum decoherence to three generations. This is rather well motivated 
 since nowadays the  atmospheric and solar neutrino 
 observations\cite{atmos,solar} can only be explained by the introduction 
 of some neutrino flavor conversion mechanism which must be understood, 
 specially after the recent impressive SNO results\cite{SNO}, in terms of 
 three generations. Therefore, this extension will be very useful, because 
 it will permit us to study the decoherence contributions on top of the 
 oscillation induced by mass (OIM) in a complete three neutrino context. 

 We have developed this three neutrino formalism using the powerful 
 technique of quantum dynamical semigroups\cite{Alicki,Davies1}, which 
 makes our analysis independent of any hypothesis about the interaction 
 between the neutrino system and the environment. 
 Quantum open systems can also be treated using the master 
 equation formalism\cite{Louisell}, but in this case, a previous knowledge 
 about the interaction with the environment is required. 
 This is an important point since it is not clearly established which is 
 the origin of the pervasive medium, the most likely possibilities at 
 the moment are quantum gravity effects described by strings at low energy 
 range. In this way, our results are broader from the phenomenological 
 point of view.
 Based on motivations coming from the master equation formalism using the 
 weak coupling limit in two generations\cite{Benatti1,Benatti2}, we have 
 casted the most relevant decoherence effects in a diagonal dissipation 
 matrix, as a result we have obtained manageable probability 
 expressions. 

 Additionally, as a phenomenological application of this three neutrino 
 extension for the decoherence phenomena, we have performed a qualitative study to inspect the agreement 
 between the average behaviour of relevant neutrino data and the 
 decoherence solution to the atmospheric neutrino anomaly in two generations, 
 embodied within this enlarged neutrino framework. We have done this study 
 assuming different alternatives for the introduction of the two generations 
 decoherence solution in our three neutrino scheme. Basically we have divided 
 the study in two parts. On the first part we use only PD to fully describe three neutrino flavor conversions, and on the second  we deal with a hybrid case, where we assume decoherence plus OIM take place in the $\nu_e \rightarrow \nu_{\mu(\tau)}$ sector.

  This paper is organized as follows. In Sec.~\ref{sec:qdn} we describe the 
 quantum dynamical semigroup formalism for a three level quantum system. 
 In Sec.~\ref{sec:neutrinos} we apply this formalism for the neutrino 
 system, explicitly calculating the survival and conversion probabilities 
 among neutrino flavors under the influence of decoherence effects.
 In Sec.~\ref{sec:pheno} we analyze the consistence of the decoherence 
 solution to the atmospheric neutrino problem in the three generation 
 framework. Finally in Sec.~\ref{sec:conclusion} we present our conclusions.

\section{Quantum dynamical semigroups and three level systems}
\label{sec:qdn}

 Hamiltonian evolution, that is, the time evolution of a physical system 
 described by the Schr\"odinger equation in the case of pure ensembles, or by 
 the Liouville equation for mixed ones, is a characteristic of systems 
 isolated from their surroundings. The time evolution of an isolated quantum 
 system with Hamiltonian $H$ is given by the continuous group of unitary 
 transformations $U_{t}=e^{iHt}$, where $t$ is the time. 
 From the mathematical point of view, the existence of the inverse of the 
 infinitesimal generator $H$, a consequence of the algebraic structure of 
 a group, gives rise to {\it reversible processes} in Hamiltonian systems.

 We know that the time evolution of a quantum open system is characterized 
 by the presence of dissipative effects which, in turn, give rise to 
 the {\it irreversible nature} of the evolution. So, if a given family of 
 transformations should be responsible for the evolution in time of a 
 quantum open system, this family will certainly not be a group. 
 A rigorous mathematical treatment of quantum open systems is provided by 
 the so called quantum dynamical semigroups\cite{Alicki,Davies1}. 

 The evolution generated by the operators in these semigroups has the 
 property of being forward in time, as a consequence of the lack of inverse 
 in a semigroup. Physically, this property can be interpreted as the 
 existence of an arrow of time which in turn makes possible the connection 
 with thermodynamics via an entropy. 

 According to Ref.\cite{Lindblad}, if  $\cal H$ is the Hilbert space of a 
 given open quantum system, and $\mathcal{B(H)}$ the space of bounded 
 operators acting on $\cal H$, the infinitesimal generator $\cal{L}$, 
 defined through its action on the density matrix $\rho(t)$, is given by      
\begin{equation}
{\mathcal L} \rho(t) = \frac{\partial \rho(t)}{\partial t} = -i[H_{\rm eff},\rho (t)] + \frac{1}{2} \sum_{j}\left([A_{j},\rho(t) A_{j}^{\dagger}] + [A_{j}\rho(t), A_{j}^{\dagger}] \right),
\label{gen}
\end{equation}
 where $H_{\rm eff}=H+H_{\rm d}$ is the ``effective'' Hamiltonian of the 
 system, $H$ being its free Hamiltonian and $H_{\rm d}$ accounts for 
 possible additional dissipative contributions which can be incorporated 
 to $H$, in other words, which can be put in the Hamiltonian form. 
 $A_j$ is a sequence of bounded operators of $\cal H$ ($A_j \in {\cal B(H)}$) 
 satisfying $\sum_{j} A_{j}^{\dagger}A_{j} \in \mathcal{B(H)}$. The first 
 term in Eq.~(\ref{gen}) constitutes the Hamiltonian part of the evolution, 
 whereas the second term is responsible for the irreversible (non-hamiltonian)
 nature of an open system evolving in time.

 Therefore, if we interpret the last expression as an effective equation 
 describing the reduced dynamics of an open system interacting with a 
 certain ``environment'', the second term in Eq.~(\ref{gen}), in a certain 
 sense, represents the interaction between the open system and the mentioned 
 environment. However, the generator $\cal L$ does not depend on a particular 
 interaction, being constructed based on very general hypothesis about the 
 time evolution, that is, irreversible dynamics, conservation of probability 
 and a less intuitive hypothesis known as complete 
 positivity\cite{Lindblad,Davies1}.      

 For a N-level system it is possible to construct an explicit 
 parameterization of Eq.~(\ref{gen}), provided that a suitable basis 
 of $\mathcal{B(H)}$, viewed as a vector space, is chosen~\cite{Sudarshan}. 
 From now on, we will restrict ourselves to a three-level system, whose 
 evolution is governed by Eq.~(\ref{gen}) and our approach will essentially 
 follow Ref.\cite{Benatti1}. The bounded operators in $\mathcal{B(H)}$ can 
 be represented by $3\times 3$ matrices of $M_{3}(\mathbf{C})$, which in 
 turn, can be generated by a basis $\{F_{\mu},\mu=0,1,...8\}$, 
 endowed with the scalar product 
 $\langle F_{\mu},F_{\nu} \rangle=\textrm{Tr}(F_{\mu}^{\dagger}F_{\nu})$ 
 and satisfying
\begin{equation}
\langle F_{\alpha},F_{\beta} \rangle = \frac{1}{2}\delta_{\alpha \beta}.
\label{scalar}
\end{equation} 

 We adopt here the standard basis of hermitian matrices 
\begin{equation}
F_{0}=\frac{1}{\sqrt{6}}{\bf 1_{3}}, \quad F_{i}=\frac{1}{2}\Lambda_{i} \quad (i=1,...,8), 
\label{basis}
\end{equation}
 where the $\Lambda_{i}$ are the Gell-Mann 
 matrices~\cite{Gell-Mann}~\footnote{There is a standard method of 
 constructing a set of matrices satisfying Eq.~(\ref{scalar}) in 
 $M_{n}(\mathbf{C})$\cite{Alicki}.}. Using this choice, $F_{i}$'s satisfy 
 the Lie algebra
\begin{equation}
\left[F_{i},F_{j}\right]=i\sum_{k}f_{ijk}F_{k}, \qquad 1\le i,j,k \le 8,
\end{equation}
 where $f_{ijk}$ are the structure constants of $SU(3)$. 
 Expanding the operators in Eq.~(\ref{gen}) in the adopted basis we 
 get~\footnote{From now on, Greek indices will range from 0 to 8, while 
 Latin indices will range from 1 to 8, unless otherwise stated.} 
\begin{equation} 
H_{\rm eff} = \sum_{\mu}h_{\mu}F_{\mu}, \quad \rho = \sum_{\mu}\rho_{\mu}F_{\mu}, \quad A_{j} = \sum_{\mu}a_{\mu}^{(j)}F_{\mu}.
\label{expansion}
\end{equation}

 As pointed out in Ref.\cite{Benatti4}, the hermiticity of $A_j$ is a 
 condition that assures the increasing with time of the von Neumman 
 entropy $S=-\textrm{Tr}(\rho \textrm{ln}\rho)$, with this choice, 
 the second term on the right hand side of Eq.~(\ref{gen}) can be written as
\begin{equation} 
\frac{1}{2}\sum_{j}\left\{[A_{j},\rho A_{j}] + [A_{j}\rho, A_{j}] \right\} = \sum_{\mu,\nu}L_{\mu \nu}\rho_{\mu}F_{\nu}, 
\label{dissipative}
\end{equation}
 where
\begin{equation}
L_{\mu0}=L_{0\mu}=0, \quad L_{ij}=\frac{1}{2}\sum_{k,l,m}\left(\vec{a}_{m} \cdot \vec{a}_{k}\right)f_{iml}f_{lkj},
\label{elements}
\end{equation}
 and in the last equation we have introduced the vectors 
 $\vec{a}_{\mu}=\{a_{\mu}^{(1)},a_{\mu}^{(2)},...,a_{\mu}^{(8)}\}$ 
 of $\mathbf{R}^{8}$ with the usual scalar product 
 $\vec{a}_{\mu} \cdot \vec{a}_{\nu}=\sum_{j}a_{\mu}^{(j)}a_{\nu}^{(j)}$. 
 $L_{\mu\nu}$ is a real~\footnote{The real nature of its entries can be 
 easily deduced from the hermiticity of $A_j$.}, symmetric matrix, defined 
 according to Eqs.~(\ref{dissipative}) and (\ref{elements}). 
 The elements of $L_{\mu\nu}$ are not totally arbitrary, but satisfy some 
 relations due to the presence of the scalar product, which in turn, should 
 obey the Cauchy-Schwartz inequality. Introducing the remaining expansions 
 of Eq.~(\ref{expansion}) into Eq.~(\ref{gen}) we get finally
\begin{equation}
\dot{\rho}_{\mu}=\sum_{i,j}h_{i}\rho_{j}f_{ij\mu} + \sum_{\nu}L_{\mu \nu}\rho_{\nu} \quad \mu = 0,...,8.
\label{differential}
\end{equation}

 The system of first order differential equations above describes the time 
 evolution of a three-level quantum open system. Such an evolution is 
 governed by the laws of quantum dynamical semigroups. The physical processes 
 exhibit an arrow of time as a consequence of the monotonical increase of the 
 von Neumman entropy as a function of time. The evolution also allows the 
 interpretation of the eigenvalues of $\rho(t)$ at any instant of time as the 
 probability of finding the open system in the eigenstate associated with 
 the eigenvalue. 

 As already mentioned, the theoretical approach provided by quantum dynamical 
 semigroups is a very general one to treat open quantum systems in the sense 
 that no explicit hypothesis has to be made about the possible interactions 
 causing the loss of quantum coherence. On the other hand, there is another 
 approach to deal with systems of the same nature, in which a well defined 
 form of the interaction ought to be provided in order to derive the 
 reduced dynamics of the open quantum system. 
 This approach is known as the master equation formalism and its rigorous 
 mathematical formulation can be found in a series of 
 papers~\cite{Gorini,Frigerio,Sudarshan,Kossakowski,Ingarden,Davies2,Martin,Spohn}.
 Furthermore, it is possible to show that the master equation formalism 
 applied to an open system weakly coupled to some environment leads to an 
 evolution equation for the reduced density matrix of the open system 
 compatible of that obtained from Eq.~(\ref{gen}) \cite{Alicki}, in 
 other words, in this situation, the master equation  and the quantum 
 dynamical semigroup formalism are equivalent.

 Coming back to Eq.~(\ref{differential}), we see that the differential 
 equation for the $\rho_{0}(t)$ component and conservation of probability 
 imply $\rho_{0}(t)=\sqrt{2/3}$. The remaining differential equations 
 have the form
\begin{equation}
\dot{\rho}_{k}=\sum_{j}\left(\sum_{i}h_{i}f_{ijk}+L_{kj}\right)\rho_{j}=\sum_{j}M_{kj}\rho_{j},
\label{diff2}
\end{equation}
 which in the matrix form can be written as \footnote{Note that 
 $\varrho \ne \rho$, the first being a column vector with 
 components $\rho_{k}$.} 
\begin{equation}
\dot{{\mathbf \varrho}}={\bf M}{\mathbf \varrho},
\label{matform}
\end{equation}
 so that the formal solution is given by  
\begin{equation}
{\mathbf \varrho}(t)=e^{{\bf M}t}{\mathbf \varrho}(0).
\label{formsol}
\end{equation}

 If $\{\lambda_{1},...,\lambda_{8}\}$ and $\{{\bf v}_{1},...,{\bf v}_{8}\}$ 
 are the set of eigenvalues and eigenvectors of $\mathbf{M}$, respectively, 
 the matrix $\mathbf{D}$, with entries $D_{ij}=(\mathbf{v}_{i})_{j}$, 
 diagonalizes $\mathbf{M}$,
\begin{equation}
{\bf M}'={\bf D}^{-1}{\bf MD}=diag(\lambda_{1},\ldots,\lambda_{8}).
\label{diagonal}
\end{equation}

 Applying the inverse transformation, 
 $e^{{\bf M}t}=\mathbf{D}e^{{\mathbf M}'t}\mathbf{D}^{-1}$, we get 
\begin{equation}
\rho_{i}(t) = \sum_{k,j}e^{\lambda_{k}t}D_{ik}D_{kj}^{-1}\rho_{j},
\end{equation}
 so that the complete solution of the system of differential equations 
 (\ref{differential}) is equivalent to a problem of eigenvalues and 
 eigenvectors. In the case of a matrix of order 8, with all non null entries, 
 the solution is too complicated to allow for direct physical 
 interpretations.

\section{Dissipative effects and neutrinos}
\label{sec:neutrinos}

 In this section we present an expression for the probability of neutrino 
 flavor conversion assuming that the dynamics responsible by this process 
 is constituted by the already known standard OIM mechanism, as well as by dissipative effects according to quantum dynamical semigroups.  

\subsection{The probability of conversion}

 The hamiltonian $H$ for a free relativistic neutrino with momentum 
 $\mathbf{p}$ and rest mass $m$  is given by
\begin{equation}
H=\sqrt{{\mathbf p}^{2}+m^{2}},
\label{hamiltonian}
\end{equation} 
 so that in the basis of mass eigenstates 
\begin{equation}
{\mathbf p}|\nu_{k}\rangle={\mathbf p}_{k}|\nu_{k}\rangle \quad \textrm{and} \quad m|\nu_{k}\rangle=m_{k}|\nu_{k}\rangle,
\end{equation}
 and we can write for relativistic neutrinos ($p=|{\mathbf p}|$),   
\begin{equation}
H \sim p^{2}+\frac{m^{2}}{2p} \Longrightarrow \langle \nu_{k}|H|\nu_{l}\rangle=\delta_{kl}\left(p_{k}+\frac{m_{k}^{2}}{2p_{k}}\right).
\end{equation} 

 The expansion of $H$ in the basis $F_{\mu}$ is
\begin{equation}
H  = \frac{1}{2p}\sqrt{\frac{2}{3}}\left(6p^{2}+\sum m^{2}\right)F_{0}+\frac{1}{2p}\left(\Delta m_{12}^{2}\right)F_{3}+\frac{1}{2\sqrt{3}p}\left(\Delta m_{13}^{2}+\Delta m_{23}^{2}\right)F_{8}, 
\label{expham}
\end{equation}
 where $\sum m^{2}=m_{1}^{2}+m_{2}^{2}+m_{3}^{2}$ and 
 $\Delta m_{ij}^{2}=m_{i}^{2}-m_{j}^{2},~i,j=1,2,3$. 

 As already mentioned at the end of the preceding section, an analysis 
 considering the most general form of $\mathbf{M}$ would certainly not provide 
 information subject to direct physical interpretation. From this point of 
 view, its reasonable to begin the analysis introducing simplifications which 
 allow a direct physical interpretation of the results. Of course, these 
 simplifications cannot be completely arbitrary, because all the physics 
 depends on them. In this way, motivations coming from the formalism of master 
 equation, in the so called weak coupling limit, applied to the problem of 
 neutrino flavor conversion in the two generation case\cite{Benatti2}, 
 lead us to assume a diagonal matrix $L_{\mu\nu}$, that is,
\begin{equation}
[L_{\mu \nu}] = diag(0,-\gamma_{1},-\gamma_{2},-\gamma_{3},-\gamma_{4},-\gamma_{5},-\gamma_{6},-\gamma_{7},-\gamma_{8}),
\label{ansatz}
\end{equation}
 where the diagonal elements are given by
\begin{equation}
\gamma_{i}=L_{ii} = -\frac{1}{2}\sum_{k,l,m}(\vec{a}_{m} \cdot \vec{a}_{k})f_{iml}f_{ikl}.
\label{diagel}
\end{equation}

 In the case of two generations, the diagonal form of $L_{\mu\nu}$ can be 
 deduced via the master equation formalism in the weak coupling limit, 
 provided that a gas of quanta, satisfying infinite statistics (for example 
 a D0-brane) is adopted as the dissipative medium. These quanta should also 
 be in thermodynamic equilibrium at a finite temperature $\beta =1/M$, where 
 $M$ defines an energy scale at which the dissipative effects are believed to 
 become important (possibly the Planck scale if the effects have quantum 
 gravity as their source). The diagonal form is finally obtained if one 
 further imposes the condition of entropy increase for finite $\beta$, for 
 details see Ref.\cite{Benatti2}. We stress that here Eq.~(\ref{ansatz}) is 
 just an {\it Ansatz} motivated by the results in the two generation case.     

 Using this parameterization of $L_{\mu\nu}$ we can construct the 
 matrix $\mathbf{M}$ defined in Eq.~(\ref{diff2}), and its eigenvalues 
 and eigenvectors can be obtained. The explicit form of $\mathbf{M}$, its 
 eigenvalues and associated eigenvectors can be found in  
 Appendix~\ref{appendixa}. 

 If at $t=0$ a neutrino of flavor $\nu_{\alpha}$ is produced, the 
 probability of its conversion, after a time $t$, to a flavor $\nu_{\beta}$ 
 can be written as    
\begin{equation}
P_{\nu_{\alpha}\rightarrow \nu_{\beta}}(t) = \textrm{Tr}[\rho^{\alpha}(t)\rho^{\beta}] = \frac{1}{3}+\frac{1}{2}\sum_{i,k,j}e^{\lambda_{k}t}D_{ik}D_{kj}^{-1}\rho_{j}^{\alpha}(0)\rho_{i}^{\beta}.
\end{equation}

 In a more explicit form, the probability of conversion can be written as
\begin{eqnarray}
& &P_{\nu_{\alpha}\rightarrow \nu_{\beta}}(t) = \frac{1}{3}+\frac{1}{2}\left\{\left[\left( \rho_{1}^{\alpha}\rho_{1}^{\beta}+\rho_{2}^{\alpha}\rho_{2}^{\beta}\right)\left(\frac{e^{-\frac{\Omega_{12}t}{2}}+e^{\frac{\Omega_{12}t}{2}}}{2}\right) \right. \right. \nonumber \\
&-&  \left. \left(\frac{ 2\Delta_{12}\left(\rho_{1}^{\alpha}\rho_{2}^{\beta}-\rho_{2}^{\alpha}\rho_{1}^{\beta}\right)+\Delta\gamma_{12}\left(\rho_{1}^{\alpha}\rho_{1}^{\beta}-\rho_{2}^{\alpha}\rho_{2}^{\beta} \right)}{\Omega_{12}}\right)\left(\frac{e^{-\frac{\Omega_{12}t}{2}}-e^{\frac{\Omega_{12}t}{2}}}{2}\right)\right]e^{-\frac{1}{2}(\gamma_{1}+\gamma_{2})t}  \nonumber \\
&+& \left[\left( \rho_{4}^{\alpha}\rho_{4}^{\beta}+\rho_{5}^{\alpha}\rho_{5}^{\beta}\right)\left(\frac{e^{-\frac{\Omega_{13}t}{2}}+e^{\frac{\Omega_{13}t}{2}}}{2}\right) \right. \nonumber \\
&-&  \left.\left(\frac{2\Delta_{13}\left( \rho_{4}^{\alpha}\rho_{5}^{\beta}-\rho_{5}^{\alpha}\rho_{4}^{\beta}\right)+\Delta\gamma_{45}\left(\rho_{4}^{\alpha}\rho_{4}^{\beta}-\rho_{5}^{\alpha}\rho_{5}^{\beta} \right)}{\Omega_{13}}\right)\left(\frac{e^{\frac{-\Omega_{13}t}{2}}-e^{\frac{\Omega_{13}t}{2}}}{2}\right)\right]e^{-\frac{1}{2}(\gamma_{4}+\gamma_{5})t} \nonumber \\
&+& \left[\left( \rho_{6}^{\alpha}\rho_{6}^{\beta}+\rho_{7}^{\alpha}\rho_{7}^{\beta}\right)\left(\frac{e^{-\frac{\Omega_{23}t}{2}}+e^{\frac{\Omega_{23}t}{2}}}{2}\right) \right. \nonumber \\
&-&  \left.\left(\frac{2\Delta_{23}\left( \rho_{6}^{\alpha}\rho_{7}^{\beta}-\rho_{7}^{\alpha}\rho_{6}^{\beta}\right)+\Delta\gamma_{67}\left(\rho_{6}^{\alpha}\rho_{7}^{\beta}-\rho_{7}^{\alpha}\rho_{7}^{\beta} \right)}{\Omega_{23}}\right)\left(\frac{e^{-\frac{\Omega_{23}t}{2}}-e^{\frac{\Omega_{23}t}{2}}}{2}\right)\right]e^{-\frac{1}{2}(\gamma_{6}+\gamma_{7})t} \nonumber \\
&+& e^{-\gamma_{3}t}\rho_{3}^{\alpha}\rho_{3}^{\beta}+e^{-\gamma_{8}t}\rho_{8}^{\alpha}\rho_{8}^{\beta} \Bigg\}\,,
\label{final}
\end{eqnarray}
 where $\Delta\gamma_{ij}=\gamma_{j}-\gamma_{i}$ and the definition of the new 
 variables $\Omega_{ij}$ can be found in Appendix~\ref{appendixa}. 

 It is easy to verify that:  Eq.~(\ref{final}) exhibits conservation of 
 probability at any instant of time; reducing the number of dimensions 
 from 3 to 2, we get the results for decoherence in two 
 generations~\cite{Benatti1}; and in the limit 
 $\gamma_{i} \rightarrow 0 \; (1\le i\le 8)$, the standard OIM mechanism is recovered.

\section{Phenomenological application}  
\label{sec:pheno}

 In this section we will use the just developed three neutrino framework 
 for OIM mechanism plus the decoherence effects, to analyze 
 the robustness of the decoherence solution to the atmospheric neutrino 
 problem in two generations~\cite{lisi} envisaged within this enlarged scheme. 
 This will be done for different selections of the mixing matrix, first for PD and second for decoherence plus masses and mixing. 
 
\subsection{Pure Decoherence} 

 In order to obtain the probabilities that correspond to the PD case it is enough to eliminate 
 the usual oscillatory terms. This can be easily done by 
 setting the mixing matrix $U$ equal to the unity matrix. Applying this   
 to Eq.~(\ref{final}) we get the following expressions: 
\begin{equation}
P^{\text{PD}}_{\nu_{\alpha} \rightarrow \nu_{\beta}}(t)=\frac{1}{3}+\frac{1}{2}(e^{-\gamma_{3}t}\rho_{3}^{\alpha}\rho_{3}^{\beta}+e^{-\gamma_{8}t}\rho_{8}^{\alpha}\rho_{8}^{\beta}),
\end{equation}
 where $\alpha,\beta=e,\mu,\tau$, so consequently
\begin{eqnarray}
 P^{\text{PD}}_{\nu_{e} \rightarrow \nu_{e}}(t)&=&\frac{1}{3}+\frac{1}{2}e^{-\gamma_{3}t}+\frac{1}{6}e^{-\gamma_{8}t}, \label{decohprob1} \\
 P^{\text{PD}}_{\nu_{e} \rightarrow \nu_{\mu}}(t)&=&\frac{1}{3}-\frac{1}{2}e^{-\gamma_{3}t}+\frac{1}{6}e^{-\gamma_{8}t}, \label{decohprob2} \\
P^{\text{PD}}_{\nu_{e} \rightarrow \nu_{\tau}}(t)&=&\frac{1}{3}-\frac{1}{3}e^{-\gamma_{8}t}, \label{decohprob3} \\
P^{\text{PD}}_{\nu_{\mu} \rightarrow \nu_{\mu}}(t)&=&\frac{1}{3}+\frac{1}{2}e^{-\gamma_{3}t}+\frac{1}{6}e^{-\gamma_{8}t}, \label{decohprob4} \\
P^{\text{PD}}_{\nu_{\mu} \rightarrow \nu_{\tau}}(t)&=&\frac{1}{3}-\frac{1}{3}e^{-\gamma_{8}t}. \label{decohprob5}
\end{eqnarray}

 Despite the fact that mixing is not included, we can observe from these 
 probabilities, that it is possible to have non-null neutrino flavor 
 conversion only taking into account the decoherence phenomena. 
 Indeed, these results were expected, since the same was obtained in  
 two generations~\cite{Benatti2}. 
    
 The atmospheric neutrino data can be interpreted well using 
 the PD phenomena in the two generation framework~\cite{lisi}. Thus, we can investigate if the pure decoherence mechanism extrapolated to
 three generations is still able to explain the current data. A qualitative 
 analysis of the behavior of the above probabilities in the present data 
 context can clearly demonstrate that this is not so. The explanation goes as 
 follows. 

 Since we aim to test the decoherence solution to the atmospheric 
 neutrino data in the three generations scheme, we will have to  
 consider the damping parameters with the dependence 
 $\gamma_{j} \rightarrow \gamma_{j} / E_\nu$,  in accordance with 
 the two generations assumption. Thus, the probabilities 
 defined in Eqs.~(\ref{decohprob1})-(\ref{decohprob5}) will be sensitive 
 only to  the ratio $L/E_\nu$ \footnote{In our formulae you can make the 
 substitution $t \leftrightarrow L$ (for c=1).} similar to vacuum case for the OIM mechanism. Here $L$ is the distance between  the source of 
 neutrinos and the detector. 
 Note also that these probabilities will not be modified by the presence 
 of matter, since there is no mixing involved.     

 The conjunction of all of these properties of the probabilities with the 
 fact that $P_{{\nu}_e \rightarrow {\nu}_{e}}=P_{{\nu}_{\mu} \rightarrow {\nu}_\mu}$, makes the three neutrino PD mechanism incompatible with the 
 actual neutrino data.
 One can exemplify this inconsistency by confronting the results of the 
 neutrino experiments CHOOZ~\cite{chooz} and K2K~\cite{k2k} with our theoretical predictions. Both experiments have similar $L/E_\nu$, but their results for 
 $P_{\overline{\nu}_e \rightarrow \overline{\nu}_e}$ and 
 $P_{\nu_\mu \rightarrow \nu_\mu}$ are not compatible with  
 our theoretical expectation. The CHOOZ experiment with 
 $\langle L/E_\nu \rangle \approx 1000/3~(\mbox{m/MeV})$ observed 
 $\langle P_{\nu_e \rightarrow \nu_e} \rangle \approx 1$~\footnote{The 
 probabilities defined in Eq.~\refe{final} satisfy  
 $P_{\overline{\nu}_{\alpha} \rightarrow \overline{\nu}_{\alpha}} \equiv 
 P_{\nu_{\alpha} \rightarrow \nu_{\alpha}}$.} and the K2K experiment with 
 $\langle L/E_\nu \rangle \approx 250/1.3 ~(\mbox{km/GeV})$ 
 has observed events compatible with $\langle P_{\nu_\mu \rightarrow \nu_\mu} 
\rangle  \approx 0.7$. Therefore the relation  
 $ P_{\nu_\mu \rightarrow \nu_\mu}= P_{\nu_e \rightarrow \nu_e}$ given 
 by our probabilities, is in contradiction with the results of CHOOZ and K2K.

 Another example that reinforces this contradiction is the multi-GeV SK data 
 sample~\cite{shiozawa} in combination with the CHOOZ constraint. In order to illustrate this fact, we consider the SK normalized $e$-like and $\mu$-like 
 event samples, which can be computed as follows 
\begin{eqnarray}
R_{\mu} &=& \langle P_{\nu_\mu \rightarrow \nu_\mu} 
\rangle + \frac{1}{r} \, \langle P_{\nu_e \rightarrow \nu_\mu} \rangle, \label{normrat1} \\
R_{e} &=& \langle P_{\nu_e \rightarrow \nu_e} 
\rangle +  r \, \langle P_{\nu_\mu \rightarrow \nu_e} \rangle, \label{normrat2}
\end{eqnarray} 
 where $R_{\mu}$ and $R_e$ are the normalized $\mu$-like and $e$-like event 
 samples defined as the ratios of the expected number of events considering 
 neutrino conversion over the expected number of events without  neutrino 
 flavor conversion, $r$ is the original proportion between the  muon and 
 electron neutrino fluxes. We can observe that for ranges of 
 $\langle L/E_{\nu} \rangle \approx 1000/3 $ compatible with CHOOZ and 
 corresponding to $\cos \theta_{\text{zenith}} \sim -0.22$~\footnote{The relation between the $\cos \theta_{\text{zenith}}$ and $L$ is given by $L=\sqrt{(R_T+h)^2-(R_T\, \sin \Theta_Z)^2}-R_T \, \cos \theta_{\text{zenith}}$, where $R_T$ is the earth radius and $h$ is the height of the neutrino production point. 
} in the multi-GeV 
 sample, $\langle E_\nu \rangle \approx 10 \, \, \mbox{GeV}$~\cite{kajita}, 
 $R_{\mu}=0.6-0.7$ and $R_e=1$. Thus, crossing this information with the 
 CHOOZ bound, we find that 
 $\langle P_{\nu_\mu \rightarrow \nu_e} \rangle = \langle P_{\nu_e \rightarrow \nu_\mu} \rangle \approx 0$ (in fact, for a real mixing matrix, 
 $ P_{\nu_{\alpha} \rightarrow \nu_{\beta}}=P_{\nu_{\beta} \rightarrow \nu_{\alpha}}$ as can be seen from Eq.~\refe{final}) and consequently 
 $\langle P_{\nu_\mu \rightarrow \nu_\mu} \rangle  \approx 0.6-0.7$, which 
 is again in conflict with our theoretical prediction.

\subsection{Decoherence $\bigoplus$ Oscillation} 

 Here we will study two different cases also admitting the presence 
 of the oscillatory terms in the probability expressions, therefore we 
 have here a mixed situation of decoherence $\bigoplus$ oscillation
 (D$\bigoplus$O). 
 In both cases the oscillation will be included through non-null mixing 
 matrix elements, which connect $\nu_e$ with $\nu_\mu$ or $\nu_e$ with 
 $\nu_\tau$. This is chosen in order to preserve the possibility to explain 
 the solar neutrino data through of the OIM mechanism. 
 In addition, another fact in our strategy of analysis  will be to adjust 
 the parameters in the $P_{\nu_\mu \rightarrow \nu_\mu}$, the leading 
 channel for the explanation of the atmospheric data, to be similar 
 to its form in the two generations decoherence solution to the 
 atmospheric neutrino problem.   

 When we introduce non-zero masses for the neutrinos, two 
 different situations may arise according to the magnitudes of 
 $\Delta_{ij}$ and $\Delta\gamma_{kl}$: either  
 $2|\Delta_{ij}| \ge |\Delta\gamma_{kl}|$ or 
 $2|\Delta_{ij}| < |\Delta\gamma_{kl}|$. In our analysis we will assume 
 the first situation, where $\Omega_{ij}$ is imaginary so that 
 Eq.~\refe{final} takes the form 

\begin{eqnarray}
& &P_{\nu_{\alpha}\rightarrow \nu_{\beta}}(t) = \frac{1}{3}+\frac{1}{2}\left\{\Bigg[\left( \rho_{1}^{\alpha}\rho_{1}^{\beta}+\rho_{2}^{\alpha}\rho_{2}^{\beta}\right)\cos\left(\frac{|\Omega_{12}|t}{2}\right) \right.  \nonumber \\
&+&  \left(\frac{ 2\Delta_{12}\left(\rho_{1}^{\alpha}\rho_{2}^{\beta}-\rho_{2}^{\alpha}\rho_{1}^{\beta}\right)+\Delta\gamma_{12}\left(\rho_{1}^{\alpha}\rho_{1}^{\beta}-\rho_{2}^{\alpha}\rho_{2}^{\beta} \right)}{\Omega_{12}}\right)\sin\left(\frac{|\Omega_{12}|t}{2}\right)\Bigg]e^{-\frac{1}{2}(\gamma_{1}+\gamma_{2})t}  \nonumber \\
&+& \Bigg[\left( \rho_{4}^{\alpha}\rho_{4}^{\beta}+\rho_{5}^{\alpha}\rho_{5}^{\beta}\right)\cos\left(\frac{|\Omega_{13}|t}{2}\right)  \nonumber \\
&+&  \left(\frac{2\Delta_{13}\left( \rho_{4}^{\alpha}\rho_{5}^{\beta}-\rho_{5}^{\alpha}\rho_{4}^{\beta}\right)+\Delta\gamma_{45}\left(\rho_{4}^{\alpha}\rho_{4}^{\beta}-\rho_{5}^{\alpha}\rho_{5}^{\beta} \right)}{\Omega_{13}}\right)\sin\left(\frac{|\Omega_{13}|t}{2}\right)\Bigg]e^{-\frac{1}{2}(\gamma_{4}+\gamma_{5})t} \nonumber \\
&+& \Bigg[\left( \rho_{6}^{\alpha}\rho_{6}^{\beta}+\rho_{7}^{\alpha}\rho_{7}^{\beta}\right) \cos\left(\frac{|\Omega_{23}|t}{2}\right) \nonumber \\
&+&  \left(\frac{2\Delta_{23}\left( \rho_{6}^{\alpha}\rho_{7}^{\beta}-\rho_{7}^{\alpha}\rho_{6}^{\beta}\right)+\Delta\gamma_{67}\left(\rho_{6}^{\alpha}\rho_{7}^{\beta}-\rho_{7}^{\alpha}\rho_{7}^{\beta} \right)}{\Omega_{23}}\right)\sin\left(\frac{|\Omega_{23}|t}{2}\right)\Bigg]e^{-\frac{1}{2}(\gamma_{6}+\gamma_{7})t} \nonumber \\
&+& e^{-\gamma_{3}t}\rho_{3}^{\alpha}\rho_{3}^{\beta}+e^{-\gamma_{8}t}\rho_{8}^{\alpha}\rho_{8}^{\beta} \Bigg\}.
\end{eqnarray}

 In the case $2|\Delta_{ij}| < |\Delta\gamma_{kl}|$, we would have an 
 analogous expression, but with hyperbolic sines and cosines. 

\subsubsection{Mixing in the $\nu_e$-$\nu_\mu$ sector}

 In this case we switch on only the non-null mixing 
 matrix elements in the $\nu_e$-$\nu_\mu$ sector, so that 
 the mixing matrix is defined as
\begin{equation}
\nonumber
U=\left(
\begin{array}{ccc}
\cos\theta & \sin\theta & 0 \\
-\sin\theta & \cos\theta & 0 \\
0 & 0 & 1
\end{array} \right).
\end{equation}

 Due to the form adopted for the mixing matrix, oscillatory terms 
 will not appear in the transition probability between $\nu_\tau$ and 
 the other flavors, remaining only the decoherence terms for generating  
 non-zero neutrino conversion related to $\nu_\tau$.

 The survival and conversion probabilities of interest for the 
 case $2|\Delta_{ij}|\ge |\Delta\gamma_{kl}|$ can be explicitly built 
 now and are given by
\newpage

 \begin{eqnarray}
 P^{\rm{\tiny{\tiny{D\oplus O}}}}_{\nu_{e} \rightarrow \nu_{e}}(t)&=&\frac{1}{3}+\frac{1}{2}\sin^{2}2\theta\left[\cos\left(\frac{|\Omega_{12}|t}{2}\right)+\frac{\Delta\gamma_{12}}{|\Omega_{12}|}\sin\left(\frac{|\Omega_{12}|t}{2}\right)\right]e^{-\frac{1}{2}(\gamma_{1}+\gamma_{2})t} \\
& & +\frac{1}{2}\cos^{2}2\theta e^{-\gamma_{3} t}+\frac{1}{6}e^{-\gamma_{8}t}, \\
P^{\rm{\tiny{\tiny{D\oplus O}}}}_{\nu_{e} \rightarrow \nu_{\mu}}(t)&=&\frac{1}{3}-\frac{1}{2}\sin^{2}2\theta\left[\cos\left(\frac{|\Omega_{12}|t}{2}\right)+\frac{\Delta\gamma_{12}}{|\Omega_{12}|}\sin\left(\frac{|\Omega_{12}|t}{2}\right)\right]e^{-\frac{1}{2}(\gamma_{1}+\gamma_{2})t} \nonumber \\
& & -\frac{1}{2}\cos^{2}2\theta e^{-\gamma_{3} t}+\frac{1}{6}e^{-\gamma_{8}t}, \\
P^{\rm{\tiny{\tiny{D\oplus O}}}}_{\nu_{e} \rightarrow \nu_{\tau}}(t)&=&\frac{1}{3}-\frac{1}{3}e^{-\gamma_{8}t}, \\
P^{\rm{\tiny{\tiny{D\oplus O}}}}_{\nu_{\mu} \rightarrow \nu_{\mu}}(t)&=&\frac{1}{3}+\frac{1}{2}\sin^{2}2\theta\left[\cos\left(\frac{|\Omega_{12}|t}{2}\right)+\frac{\Delta\gamma_{12}}{|\Omega_{12}|}\sin\left(\frac{|\Omega_{12}|t}{2}\right)\right]e^{-\frac{1}{2}(\gamma_{1}+\gamma_{2})t} \nonumber \\
& & +\frac{1}{2}\cos^{2}2\theta e^{-\gamma_{3} t}+\frac{1}{6}e^{-\gamma_{8}t}, \\
P^{\rm{\tiny{\tiny{D\oplus O}}}}_{\nu_{\mu} \rightarrow \nu_{\tau}}(t)&=&\frac{1}{3}-\frac{1}{3}e^{-\gamma_{8}t},  
\end{eqnarray}
 and analogous expressions can be written for the 
 case $2|\Delta_{ij}| < |\Delta\gamma_{kl}|$, through the 
 substitutions $\cos \rightarrow \cosh$, $\sin \rightarrow \sinh$. 

 Once we have defined these probabilities we will check their 
 consistency with the tendency indicated by the atmospheric neutrino 
 data. Our test  will be based upon the definitions for the 
 normalized $e$-events and $\mu$-events given in the 
 Eqs.~\refe{normrat1} and \refe{normrat2}.  

 Since our probabilities involving  mixing with $\nu_e$ are 
 going to be applied for the atmospheric neutrino data, matter effects 
 must be considered. This is because the atmospheric neutrino data involve 
 neutrinos which travel different distances through the Earth. 
 The inclusion of matter effects in the probability expressions is 
 really straightforward in the case where constant matter density is 
 considered~\cite{Benatti2}. Basically, we only need to replace 
 $\theta \rightarrow \theta_{m}$ (in the corresponding sines and cosines), 
 and  $\Delta_{13} \rightarrow \Delta^{m}_{13}$ where 
\begin {equation}
\sin^2 2  \theta_m (x) = 
{ \sin^2 2 \theta \over 
\sin^2 2 \theta  +  (x - \cos 2 \theta)^2 },
\label {eq:st2}
\end{equation}
and, 
 \begin{equation} 
\Delta^{m}_{13}= \sqrt{ \sin^2 2 \theta  +  (x - \cos 2 \theta)^2} \Delta_{13},
\label{eq:mm} 
\end{equation} 
where $x$ is given by
\begin{equation}     
x=\frac{2 \sqrt{2} \, G_{F} \, n_{e} \, E_{\nu}} 
{\Delta m^{2}_{13}}\simeq 0.76 \; {\rho(\mbox{g}\,\mbox{cm}^{-3}) 
~E_\nu(\mbox{GeV}) \over 
\Delta m^2_{13} (10^{-4}~\mbox{eV}^2) },
\end{equation}
 $G_{F}$ is the Fermi constant, $n_{e}$ the electron number  
 density, $E_{\nu}$ the neutrino energy and 
 $\rho$ the matter density.

 For evaluating how these matter effects affect our probabilities, we will 
 choose the following values for the relevant parameters: 
 ${\Delta m^{2}_{13}} \leq  3 \times 10^{-4}~\mbox{eV}^2$, consistent 
 with the large mixing angle (LMA) solution to the solar neutrino problem~\cite{smy}, $E_\nu \sim 10 \mbox{ GeV}$, consistent with the mean neutrino energy for multi-GeV events, and $\rho(\mbox{g}\,\mbox{cm}^{-3}) \simeq 2.75$~\cite{prem}, this value is one of the lowest values in the Earth matter density profile. 
 Using these values we estimate that $\sin^2 2  \theta_m \rightarrow 0$ and 
 $\cos^2 2  \theta_m \rightarrow 1$. Note that this will happen even 
 faster for higher values of matter density. As a result, we can write 
 simplified probabilities for this case       
\begin{eqnarray}
P^{\rm{\tiny{\tiny{D\oplus O}}}}_{\nu_{e} \rightarrow \nu_{e}}(t)& \simeq &\frac{1}{3}+\ \frac{1}{2} e^{-\gamma_{3} t}+\frac{1}{6}e^{-\gamma_{8}t},\\ 
P^{\rm{\tiny{\tiny{D\oplus O}}}}_{\nu_{\mu} \rightarrow \nu_{e}}(t)& \simeq &\frac{1}{3} -\frac{1}{2} e^{-\gamma_{3} t}+\frac{1}{6}e^{-\gamma_{8}t},\\ 
P^{\rm{\tiny{\tiny{D\oplus O}}}}_{\nu_{\mu} \rightarrow \nu_{\mu}}(t)& \simeq &\frac{1}{3}+ \frac{1}{2} e^{-\gamma_{3} t}+\frac{1}{6}e^{-\gamma_{8}t},
\end{eqnarray}
 here we go one step further and assume $\gamma_{8}t \rightarrow 0$,  
\begin{equation}
P^{\rm{\tiny{\tiny{D\oplus O}}}}_{\nu_{\mu} \rightarrow \nu_{\mu}}(t)=\frac{1}{2}+\frac{1}{2}e^{-\gamma_{3}t},
\end{equation}
 thus we mimic completely the two generation expression 
 for decoherence in $P_{\nu_{\mu} \rightarrow \nu_{\mu}}$.

 At this point we have $P_{\nu_{\mu} \rightarrow \nu_{\mu}}=P_{\nu_{e} \rightarrow \nu_{e}}=P$, $P_{\nu_{e} \rightarrow \nu_{\mu}}=1-P$ 
and $P_{\nu_{e} \rightarrow \nu_{\tau}}=P_{\nu_{\mu} \rightarrow 
\nu_{\tau}}=0$. 
 Writing Eqs.~\refe{normrat1} and \refe{normrat2} as a function of $P$, 
 we obtain 
\begin{eqnarray}
R_{\mu} &=& \left( 1-\frac{1}{r} \right) \langle P \rangle + \frac{1}{r}, \\ 
R_{e} &=& \left( 1-r \right) \langle P \rangle +  r.  
\end{eqnarray}

 Now we can check, in a model independent way, the consistency of these 
 expressions with the atmospheric neutrino observations. To make an 
 analysis  consistent with  our assumption on the neutrino energy, we will 
 look at the tendencies of multi-GeV events, this implies $r \approx 3$. 
 The $e$-like multi-GeV events are consistent with $R_{e} \simeq 1$ for 
 any neutrino trajectory, so that  
\begin{equation} 
R_{e} = -2 \langle P \rangle + 3 \simeq 1 \, \Rightarrow \, 
\langle P \rangle \simeq 1,    
\end{equation}
and consequently,
\begin{equation} 
R_{\mu} = \frac{2}{3} \langle P \rangle + \frac{1}{3} \simeq 1,  
\end{equation} 
 which is in strong disagreement with the behavior of neutrinos coming from 
 below the horizon, since $\mu$-like multi-GeV events are indicating in 
 average that $R_{\mu} \simeq 0.5-0.6$. Therefore this case is also in 
 conflict with the present data. 

\subsubsection{Mixing in the $\nu_e$-$\nu_\tau$ sector}

 In this case we turn on only the mixing matrix elements that 
 connect $\nu_e$-$\nu_\tau$, which implies the following mixing matrix 
\begin{equation}
\nonumber
U=\left(
\begin{array}{ccc}
\cos\theta & 0 &\sin\theta \\
0 & 1 & 0 \\
-\sin\theta & 0 & \cos\theta  \\
\end{array} \right),
\end{equation}      
 so that the mass induced oscillation terms do not appear in any 
 transition involving $\nu_\mu$. 

 Then, for the case $2|\Delta_{ij}|\ge |\Delta\gamma_{kl}|$, we obtain 
 the probability formulae
\begin{eqnarray}
P^{\rm{\tiny{\tiny{D\oplus O}}}}_{\nu_{e} \rightarrow \nu_{e}}(t)&=&\frac{1}{3}+\frac{1}{2}\cos^4\theta e^{-\gamma_{3}t}+\frac{1}{6}(\cos^2\theta-2\sin^2\theta)^{2}e^{-\gamma_{8}t}  \\
& &+\frac{1}{2}\sin^{2} 2\theta\left[\cos\left(\frac{|\Omega_{13}|t}{2}\right)+\frac{\Delta\gamma_{45}}{|\Omega_{13}|}\sin\left(\frac{|\Omega_{13}|t}{2}\right)\right] e^{-\frac{1}{2}(\gamma_{4}+\gamma_{5})t}, \\
P^{\rm{\tiny{\tiny{D\oplus O}}}}_{\nu_{e} \rightarrow \nu_{\mu}}(t)&=&\frac{1}{3}-\frac{1}{2}\cos^2\theta e^{-\gamma_{3}t}+\frac{1}{6}(\cos^2\theta-2\sin^2\theta)e^{-\gamma_{8}t},  \\
P^{\rm{\tiny{\tiny{D\oplus O}}}}_{\nu_{e} \rightarrow \nu_{\tau}}(t)&=&\frac{1}{3}+\frac{1}{2}\cos^2\theta\sin^{2}\theta e^{-\gamma_{3}t}+\frac{1}{6}(\cos^2\theta-2\sin^2\theta)(\sin^2\theta-2\cos^2\theta)e^{-\gamma_{8}t} \\
& &-\frac{1}{2}\sin^{2} 2\theta\left[\cos\left(\frac{|\Omega_{13}|t}{2}\right)+\frac{\Delta\gamma_{45}}{|\Omega_{13}|}\sin\left(\frac{|\Omega_{13}|t}{2}\right)\right]e^{-\frac{1}{2}(\gamma_{4}+\gamma_{5})t}, \\
P^{\rm{\tiny{\tiny{D\oplus O}}}}_{\nu_{\mu} \rightarrow \nu_{\mu}}(t)&=&\frac{1}{3}+\frac{1}{2}e^{-\gamma_{3}t}+\frac{1}{6}e^{-\gamma_{8}t}, \\
P^{\rm{\tiny{\tiny{D\oplus O}}}}_{\nu_{\mu} \rightarrow \nu_{\tau}}(t)&=&\frac{1}{3}-\frac{1}{2}\sin^{2}\theta e^{-\gamma_{3}t}+\frac{1}{6}(\sin^2\theta-2\cos^2\theta)^{2}e^{-\gamma_{8}t}, 
\end{eqnarray}
 and once again, the case $2|\Delta_{ij}|<|\Delta\gamma_{kl}|$ can be obtained 
 just by substituting the harmonic functions by their hyperbolic partners.

 Before analyzing the probabilities in this case, we will make some 
 further assumptions. 
 First, we will take $3 \times 10^{-5} \lesssim \Delta m^2_{13}/\mbox{eV}^{2} 
 \lesssim 19  \times 10^{-5}$
 and $0.25 \lesssim \tan^2 \theta \lesssim 0.65$, both consistent with 
 the LMA solution 
 to the solar neutrino problem~\cite{smy}, second we will neglect all the decoherence 
 parameters for the $L/E_\nu$ at the range of the atmospheric neutrino scale, 
 with the exception of $\gamma_3$, which is needed in order to make   
 $P_{\nu_\mu \rightarrow \nu_\mu}$ similar to its form in the decoherence 
 solution  to the atmospheric neutrino anomaly in  two generations. 
 
 Once we have accepted these assumptions, we can study the compatibility of 
 these probabilities with the current data. We will not use the atmospheric 
 neutrino data to test the viability of the decoherence solution in the 
 three generations scheme. This is because the probabilities 
 are written as a function of $\cos \theta$ and  $\sin \theta$, which makes 
 it difficult to extract correct conclusions without introducing    
 refinements to the qualitative analysis we have used up to now. 
 Instead, we will simply look at CHOOZ data. This will be enough to give 
 us a good idea about the compatibility of the decoherence solution in  
 the three neutrino scheme, through the observations in the  channel 
 $\nu_e \rightarrow \nu_e$. 
 Since we are going to use for this study $P_{\nu_e \rightarrow \nu_e}$, 
 it is convenient to write it down after the assumptions mentioned above are 
 taken into account  
\begin{equation}
P^{\rm{\tiny{\tiny{D\oplus O}}}}_{\nu_{e} \rightarrow \nu_{e}}(L)=\frac{1}{3}+\frac{1}{2}\left(\cos^4\theta e^{-\gamma^{\ast}_{3}L/E_{\nu}}+\sin^{2} 2\theta\cos(\Delta_{13}L) \right)+\frac{1}{6}(\cos^2\theta-2\sin^2\theta)^{2}. 
\end{equation}    

 We have obtained excluded regions in the plane 
 ($\cos^2 \theta, \gamma^{\ast}_{3}$), with 
 $\gamma_{3}=\gamma^{\ast}_{3}/E_{\nu}$, following the statistical 
 procedure described in Ref.~\cite{gstz1}.
 We must point out that variations in $\Delta m^2_{13}$ will not affect 
 these excluded regions, since $\cos(\Delta_{13}L) \rightarrow 1$ in the 
 CHOOZ range. 
 
\begin{figure}
\begin{center}
\vglue -1.5cm
\epsfig{file=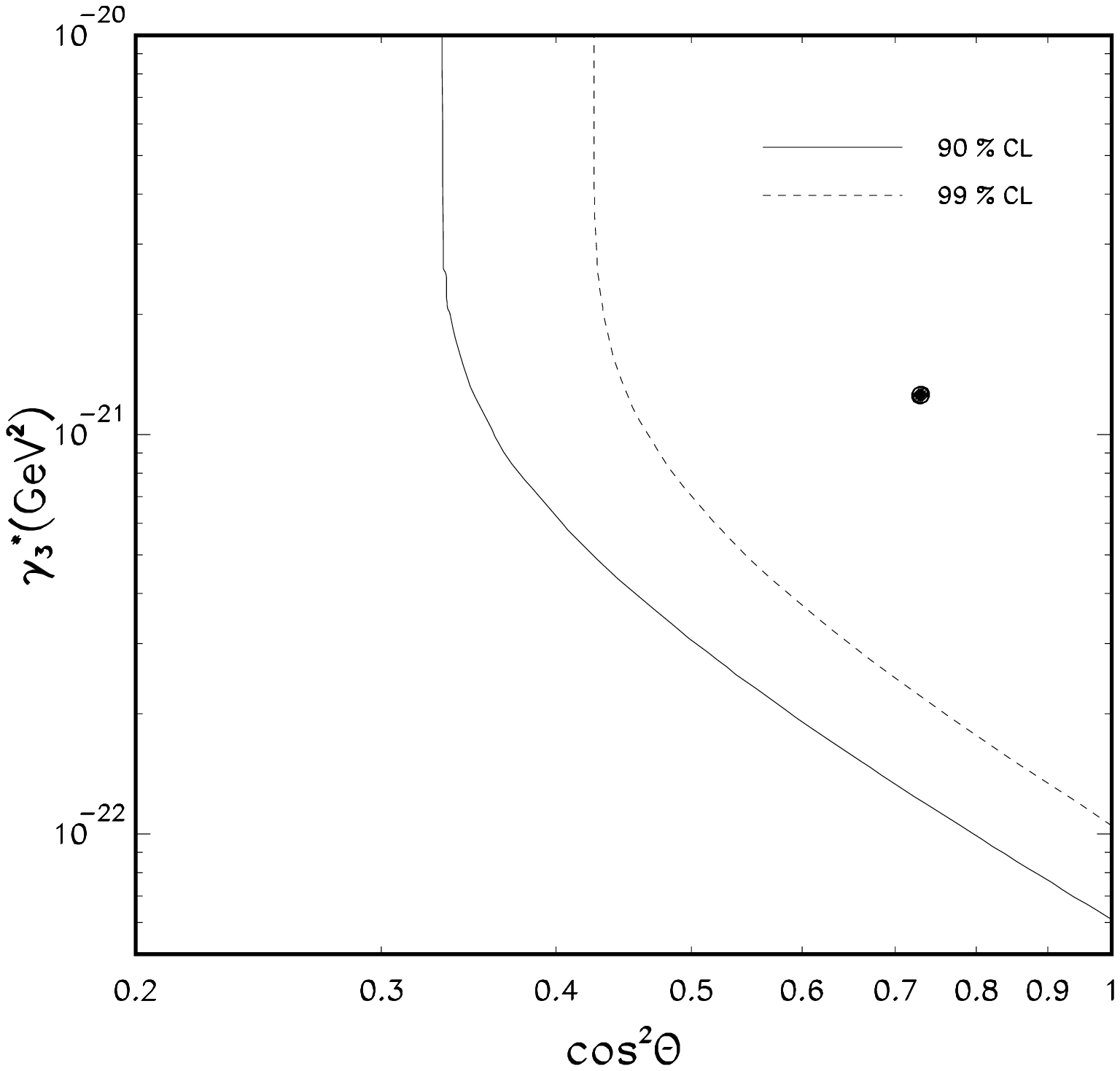,height=16.0cm,width=14.54cm}
\vglue -0.7cm
\caption{Regions in the plane $(\cos^2 \theta, \gamma_{3}^{\ast})$
 excluded by CHOOZ data. The black point denotes the best 
 fit value for the decoherence solution to the atmospheric neutrino problem, 
 $\gamma_{3}^{\ast}= 1.2 \times 10^{21}$ GeV$^2$ as well as for the 
 LMA solution to the solar neutrino one, $\cos^2 \theta \approx 0.73$.
 All points to the right of the curves are excluded at 90 \% (continuous line) 
 and 99 \% (dashed line) C.L.}
\label{fig1}
\vglue -0.05cm
\end{center}
\end{figure}

 The results are shown in Fig.~\ref{fig1}, where the points to the 
 right of the contours are excluded. We observe that the point 
 in this plane which corresponds to the value of the decoherence parameter 
 which best explains the atmospheric  neutrino data, 
 $\gamma^{\ast}_{3}=1.2 \times 10^{-21}$ 
 GeV$^{2}$, as well as $\cos^2 \theta$ which corresponds 
 to the best fit value for the LMA solution to the solar neutrino problem,
 $\cos^2 \theta \approx 0.73$ ($\sin^2 2\theta \approx 0.8$), 
 denoted by a black dot, is excluded at 
 99 \% C.L.  In general, if $\cos^2 \theta \gtrsim 0.5$ we can exclude 
 $\gamma^{\ast}_{3} \lesssim 7.0 \times 10^{-22}$ GeV$^2$
 at 99 \% C.L. 
 In this way we observe that CHOOZ data also highly 
 disfavor this three neutrino scheme, since it excludes a large region 
 of $\gamma^{\ast}_{3}$ compatible with the atmospheric neutrino solution 
 for values of $\cos^2 \theta$ consistent with the LMA solution to the solar 
 neutrino deficit. 

\section{Conclusions}
\label{sec:conclusion}

 Under the assumption that the neutrino system can interact with a 
 pervasive environment, we have obtained neutrino probability formulae 
 for three neutrino generations, taking into account quantum dissipative 
 effects coming from the interaction with the medium on top of the OIM mechanism. The damping terms were brought in 
 through the quantum dynamical semigroup formalism. This approach is 
 very useful, since no {\it a priori} assumption on the form of 
 neutrino-medium interaction has to be made. Some simplifications of the 
 form of the dissipative matrix were adopted based on results in two 
 generations.

 We have performed a qualitative analysis to test if the two generation 
 decoherence solution to the atmospheric neutrino problem viewed in this 
 hybrid three neutrino framework, can still explain the tendencies of the 
 current experimental neutrino data. 

 We have analyzed two different cases, the first one considering only PD and the second including a mixture of both conversion mechanisms,
 that is, decoherence plus OIM. The second case was further subdivided into two cases, according to the choice
 of mixing matrix: i) mass and mixing  contributions connecting 
 $\nu_e \rightarrow \nu_\mu$; ii)  mass and mixing contribution 
 connecting $\nu_e \rightarrow \nu_\tau$.  
 We have observed that all of these cases are clearly disfavored by 
 recent relevant experimental neutrino data. Particularly, in the PD case, the fact that 
 $P_{\nu_e \rightarrow \nu_e}=P_{\nu_\mu \rightarrow \nu_\mu}$ is not 
 compatible with the constraint given by CHOOZ 
 combined with SK data  ($R_e$ and $R_\mu$) or with K2K results. 
 For the hybrid case of decoherence plus non-null mixing in 
 $\nu_e \rightarrow \nu_\mu$, the same prediction 
 $P_{\nu_e \rightarrow \nu_e}=P_{\nu_\mu \rightarrow \nu_\mu}$ arises which 
 is  clearly not supported by SK data. 
 In the case of decoherence plus mixing in $\nu_e \rightarrow \nu_\tau$, 
 we have made a statistical analysis of CHOOZ data using our theoretical 
 expression for  $P_{\nu_e \rightarrow \nu_e}$, with some simplifications.
 We have obtained that the best fit value for the decoherence solution to 
 the atmospheric neutrino problem, 
 $\gamma^{\ast}_3=1.2 \times 10^{-21}$ GeV$^2$, is highly disfavored by 
 data.  Values of $\gamma^{\ast}_3 > 3. \times 10^{-22}$ GeV$^2$ for 
 $\cos \theta$ consistent with the LMA solution to the solar neutrino deficit 
 are, in general, excluded at 99 \% C.L..       

 Although, the tests we have performed in the three neutrino scheme 
 indicate a disagreement between data and theoretical expectations, this does 
 not mean that dissipative effects can not exist as subleading processes, 
 with the full three neutrino OIM as the main mechanism for neutrino flavor conversion. In fact, the formulae 
 developed here are interesting to be used to help establishing  limits 
 on the  decoherence parameters or to try to detect their effects using an 
 appropriate three neutrino description.    

 Also it is worth to stress that we have worked here in a simplified situation 
 in which the dissipative matrix is diagonal. The presence of off-diagonal 
 terms can certainly produce interesting effects on the probability 
 expressions. Our formulae can be easily modified to include these 
 off-diagonal terms. However, any further qualitative or quantitative 
 analysis in three generations will not be so direct.


\acknowledgments

 This work was supported by Conselho Nacional de Desenvolvimento 
 Cient\'{\i}fico e Tecnol\'ogico (CNPq) and by Funda\c{c}\~ao de 
 Amparo \`a Pesquisa do Estado de S\~ao Paulo (FAPESP).



\newpage


\appendix
\section{Explicit form of $\mathbf{M}$}
\label{appendixa}

We present here the explicit form of the matrix $\mathbf{M}$, its eigenvalues 
and eigenvectors, as well as the diagonalizing matrix $\mathbf{D}$ and its 
inverse defined in Eq.~(\ref{diagonal}). Starting from the definition of
$\mathbf{M}$ in Eq.~(\ref{diff2}), taking as valid the approximation 
$H_{\rm eff}\sim H$\cite{Benatti2}, we can write
\begin{equation}
{\bf M} = \left(
\begin{array}{cccccccc}
-\gamma_{1} & -\Delta_{12} & 0  & 0 & 0 & 0 & 0 & 0 \\
\Delta_{12} & -\gamma_{2}  & 0 & 0 & 0 & 0 & 0 & 0 \\
0 & 0 & -\gamma_{3}  & 0 & 0 & 0 & 0 & 0 \\
0 & 0 & 0 & -\gamma_{4}  & -\Delta_{13} & 0 & 0 & 0 \\
0 & 0 & 0 & \Delta_{13} & -\gamma_{5}  & 0 & 0 & 0 \\
0 & 0 & 0 & 0 & 0 & -\gamma_{6}  & -\Delta_{23} & 0 \\
0 & 0 & 0 & 0 & 0 & \Delta_{23} & -\gamma_{7}  & 0 \\
0 & 0 & 0 & 0 & 0 & 0 & 0 & -\gamma_{8}  \\
\end{array}\right),
\end{equation}
with the definitions
\begin{equation}
h_{3}=\Delta_{12}, \quad h_{8}=\frac{1}{\sqrt{3}}(\Delta_{13}+\Delta_{23}) \quad \textrm{and} \quad \Delta_{ij}=\frac{\Delta m_{ij}^{2}}{2p}, ~i,j=1,2,3.  
\end{equation}

Now solving the secular equation 
$\textrm{det}({\mathbf M}-\lambda {\mathbf 1})=0$, we get
\begin{eqnarray}
\lambda_{1}&=&\frac{1}{2}\left[-(\gamma_{1}+\gamma_{2})-\sqrt{(\gamma_{2}-\gamma_{1})^{2}-4\Delta_{12}^{2}}\right]=\frac{1}{2}\left[-(\gamma_{1}+\gamma_{2})-\Omega_{12}\right] \nonumber \\
\lambda_{2}&=&\frac{1}{2}\left[-(\gamma_{1}+\gamma_{2})+\sqrt{(\gamma_{2}-\gamma_{1})^{2}-4\Delta_{12}^{2}}\right]=\frac{1}{2}\left[-(\gamma_{1}+\gamma_{2})+\Omega_{12}\right] \nonumber \\
\lambda_{3}&=&-\gamma_{3} \nonumber \\
\lambda_{4}&=&\frac{1}{2}\left[-(\gamma_{4}+\gamma_{5})-\sqrt{(\gamma_{5}-\gamma_{4})^{2}-4\Delta_{13}^{2}}\right]=\frac{1}{2}\left[-(\gamma_{4}+\gamma_{5})-\Omega_{13}\right] \\
\lambda_{5}&=&\frac{1}{2}\left[-(\gamma_{4}+\gamma_{5})+\sqrt{(\gamma_{5}-\gamma_{4})^{2}-4\Delta_{13}^{2}}\right]=\frac{1}{2}\left[-(\gamma_{4}+\gamma_{5})+\Omega_{13}\right] \nonumber \\
\lambda_{6}&=&\frac{1}{2}\left[-(\gamma_{6}+\gamma_{7})-\sqrt{(\gamma_{7}-\gamma_{6})^{2}-4\Delta_{23}^{2}}\right]=\frac{1}{2}\left[-(\gamma_{6}+\gamma_{7})-\Omega_{23}\right] \nonumber \\
\lambda_{7}&=&\frac{1}{2}\left[-(\gamma_{7}+\gamma_{8})+\sqrt{(\gamma_{7}-\gamma_{6})^{2}-4\Delta_{23}^{2}}\right]=\frac{1}{2}\left[-(\gamma_{6}+\gamma_{7})+\Omega_{23}\right] \nonumber \\
\lambda_{8}&=&-\gamma_{8} \nonumber
\end{eqnarray}

The associated eigenvectors are in turn
\begin{eqnarray}
{\bf v}_{1}^{T}&=&\left(\frac{\lambda_{1}+\gamma_{2}}{\Delta_{12}},1,0,0,0,0,0,0\right) \nonumber \\
{\bf v}_{2}^{T}&=&\left(\frac{\lambda_{2}+\gamma_{2}}{\Delta_{12}},1,0,0,0,0,0,0\right) \nonumber \\ 
{\bf v}_{3}^{T}&=&\left(0,0,1,0,0,0,0,0\right)\nonumber \\
{\bf v}_{4}^{T}&=&\left(0,0,0,\frac{\lambda_{4}+\gamma_{5}}{\Delta_{13}},1,0,0,0\right)  \\
{\bf v}_{5}^{T}&=&\left(0,0,0,\frac{\lambda_{4}+\gamma_{5}}{\Delta_{13}},1,0,0,0\right)\nonumber \\
{\bf v}_{6}^{T}&=&\left(0,0,0,0,0,\frac{\lambda_{6}+\gamma_{7}}{\Delta_{23}},1,0\right)\nonumber \\
{\bf v}_{7}^{T}&=&\left(0,0,0,0,0,\frac{\lambda_{7}+\gamma_{7}}{\Delta_{23}},1,0\right)\nonumber \\
{\bf v}_{8}^{T}&=&\left(0,0,0,0,0,0,0,1\right),\nonumber 
\end{eqnarray}
where the superscript $T$ denotes transposition. One can now construct 
the diagonalizing matrix $\mathbf{D}$ and its inverse 
\begin{equation}
{\bf D}=\left(
\begin{array}{cccccccc}
\frac{\lambda_{1}+\gamma_{2}}{\Delta_{12}} & \frac{\lambda_{2}+\gamma_{2}}{\Delta_{12}} & 0 & 0 & 0 & 0 & 0 & 0 \\
1 & 1 & 0 & 0 & 0 & 0 & 0 & 0 \\
0 & 0 & 1 & 0 & 0 & 0 & 0 & 0 \\
0 & 0 & 0 & \frac{\lambda_{4}+\gamma_{5}}{\Delta_{13}} & \frac{\lambda_{5}+\gamma_{5}}{\Delta_{13}} & 0 & 0 & 0 \\
0 & 0 & 0 & 1 & 1 & 0 & 0 & 0 \\
0 & 0 & 0 & 0 & 0 & \frac{\lambda_{6}+\gamma_{7}}{\Delta_{23}} & \frac{\lambda_{7}+\gamma_{7}}{\Delta_{23}} & 0 \\
0 & 0 & 0 & 0 & 0 & 1 & 1 & 0 \\
0 & 0 & 0 & 0 & 0 & 0 & 0 & 1
\end{array}\right),
\end{equation}

\begin{equation} 
{\bf D}^{-1}=\left(
\begin{array}{cccccccc}
-\frac{\Delta_{12}}{\Omega_{12}} & \frac{\lambda_{2}+\gamma_{2}}{\Omega_{12}} & 0 & 0 & 0 & 0 & 0 & 0 \\
\frac{\Delta_{12}}{\Omega_{12}} & -\frac{\lambda_{1}+\gamma_{2}}{\Omega_{12}} & 0 & 0 & 0 & 0 & 0 & 0 \\
0 & 0 & 1 & 0 & 0 & 0 & 0 & 0 \\
0 & 0 & 0 & -\frac{\Delta_{13}}{\Omega_{13}} & \frac{\lambda_{5}+\gamma_{5}}{\Omega_{13}}  & 0 & 0 & 0 \\
0 & 0 & 0 & \frac{\Delta_{13}}{\Omega_{13}}  & -\frac{\lambda_{4}+\gamma_{5}}{\Omega_{13}} & 0 & 0 & 0 \\
0 & 0 & 0 & 0 & 0 & -\frac{\Delta_{23}}{\Omega_{23}}  & \frac{\lambda_{7}+\gamma_{7}}{\Omega_{23}}  & 0 \\
0 & 0 & 0 & 0 & 0 & \frac{\Delta_{23}}{\Omega_{23}} & -\frac{\lambda_{6}+\gamma_{7}}{\Omega_{23}} & 0 \\
0 & 0 & 0 & 0 & 0 & 0 & 0  & 1 
\end{array}\right).
\end{equation}

\newpage

\section{Explicit form of the coefficients $\rho_\mu^\alpha$}
\label{appendixb}

Flavor eigenstates $|\nu_{\alpha}\rangle, \, \alpha=e ,\mu ,\tau$ can be 
written as a function of mass eigenstates  $|\nu_{k}\rangle, \, k=1,2,3$ 
through a unitary matrix $U$
\begin{equation}
|\nu_{\alpha}\rangle=\sum_{k=1}^{3}U_{\alpha k}^{\ast}|\nu_{k}\rangle \quad \textrm{and} \quad
\sum_{k=1}^{3}U_{k \beta}U_{k \alpha}^{\ast}=\sum_{k=1}^{3}U_{\beta k}U_{\alpha k}^{\ast}=\delta_{\alpha \beta},
\end{equation}
we can express the density matrix of a flavor state  $|\nu_{\alpha}\rangle$ as
\begin{eqnarray*}
\rho^{\alpha}=|\nu_{\alpha}\rangle\langle\nu_{\alpha}| & = & \left(\sum_{k=1}^{3}U_{\alpha k}^{\ast}|\nu_{k}\rangle\right)\left(\sum_{k=1}^{3}U_{\alpha k}\langle\nu_{k}|\right) \\
& = & \sum_{k,l=1}^{3}U_{\alpha k}^{\ast}U_{\alpha l}|\nu_{k}\rangle\langle\nu_{l}|.
\end{eqnarray*}

In the mass eigenstates basis we have
\begin{equation}
\langle\nu_{m}|\rho^{\alpha}|\nu_{n}\rangle=\sum_{k,l=1}^{3}U_{\alpha k}^{\ast}U_{\alpha l}\langle\nu_{m}|\nu_{k}\rangle\langle\nu_{l}|\nu_{n}\rangle=U_{\alpha m}^{\ast}U_{\alpha n},
\end{equation}
so that
\begin{equation}
[\rho^{\alpha}]=\left(
\begin{array}{ccc}
|U_{\alpha 1}|^2 & U_{\alpha 1}^{\ast}U_{\alpha 2} & U_{\alpha 1}^{\ast}U_{\alpha 3} \\
U_{\alpha 2}^{\ast}U_{\alpha 1} & |U_{\alpha 2}|^2 & U_{\alpha 2}^{\ast}U_{\alpha 3} \\
U_{\alpha 3}^{\ast}U_{\alpha 1} & U_{\alpha 3}^{\ast}U_{\alpha 2} & |U_{\alpha 3}|^2
\end{array}\right).
\end{equation}

Therefore, the coefficients 
$\rho^{\alpha}_{\mu}=2\textrm{Tr}[\rho^{\alpha}F_{\mu}]$ can be explicitly 
written as
\begin{eqnarray}
& & \rho_{0}^{\alpha}=\sqrt{2/3} \nonumber \\
& & \rho_{1}^{\alpha}=2\;\re(U_{\alpha 1}^{\ast}U_{\alpha 2}) \nonumber \\
& & \rho_{2}^{\alpha}=-2\;\im(U_{\alpha 1}^{\ast}U_{\alpha 2})\nonumber \\
& & \rho_{3}^{\alpha}=|U_{\alpha 1}|^{2}-|U_{\alpha 2}|^{2} \nonumber \\
& & \rho_{4}^{\alpha}=2\;\re(U_{\alpha 1}^{\ast}U_{\alpha 3}) \\
& & \rho_{5}^{\alpha}=-2\;\im(U_{\alpha 1}^{\ast}U_{\alpha 3}) \nonumber \\
& & \rho_{6}^{\alpha}=2\;\re(U_{\alpha 2}^{\ast}U_{\alpha 3}) \nonumber \\
& & \rho_{7}^{\alpha}=-2\;\im(U_{\alpha 2}^{\ast}U_{\alpha 3}) \nonumber \\
& & \rho_{8}^{\alpha}=\frac{1}{\sqrt{3}}(|U_{\alpha 1}|^{2}+|U_{\alpha 2}|^{2}-2|U_{\alpha 3}|^{2}) \nonumber
\end{eqnarray}


\begin{thebibliography}{99}

\bibitem{Davies1} E. B. Davies, {\it Quantum Theory of Open Systems}, 
 Academic Press, London (1976).

\bibitem{Louisell} W. H. Louisell, {\it Quantum Statistical Properties of 
Radiation}, Jonh Wiley \& Sons, New York (1973). 

\bibitem{Benatti1} F. Benatti and R. Floreanini, JHEP {\bf 0002}, 32 (2000).

\bibitem{Benatti2} F. Benatti and R. Floreanini, {\it Phys. Rev. D} 
 {\bf 64}, 085015 (2001). 

\bibitem{gstz1} A. M. Gago, E. M. Santos, W. J. C. Teves and R. Zukanovich 
Funchal, {\it Phys. Rev. D} {\bf 63}, 073001 (2001).

\bibitem{gstz2} A. M. Gago, E. M. Santos, W. J. C. Teves and R. Zukanovich 
Funchal, {\it Phys. Rev. D} {\bf 63}, 113013 (2001).

\bibitem{lisi} E. Lisi, A. Marrone and, D. Montanino, {\it Phys. Rev. Lett.} {\bf 85}, 1166 (2000).

\bibitem{atmos} Kamiokande Collaboration, H. S. Hirata  {\it et al.},
{\it Phys. Lett.} {\bf B 205}, 416 (1988); {\it ibid.} {\bf 280}, 
 146 (1992); Y. Fukuda {\it et al.}, {\it ibid.}{\bf 335}, 237 (1994); 
 IMB Collaboration, R. Becker-Szendy  {\it et al.}, {\it Phys. Rev. D} 
 {\bf 46}, 3720 (1992); 
 Soudan-2 Collaboration, W. W. M. Allison  {\it et al.}, {\it Phys. Lett.} 
 {\bf B 391}, 491 (1997); Super-Kamiokande Collaboration, 
 Y. Fukuda {\it et al.}, {\it Phys. Rev. Lett.} {\bf 81}, 1562 (1998);
 {\it Phys. Lett.} {\bf B 436}, 33 (1999).

\bibitem{solar} Homestake Collaboration, K.\ Lande  {\it et al.}, 
{\it Astrophys. J.} {\bf 496}, 505 (1998); {\it Nucl. Phys.} {\bf 77} 
 ({\it Proc. Suppl.}), 13 (1999); Gallex Collaboration,  
 W. Hampel {\it et al.}, {\it Phys. Lett.} {\bf B 447}, 127 (1999); 
 GNO Collaboration,  M. Altmann  {\it et al.}, {\it Phys. Lett.} 
 {\bf B 490}, 16 (2000); SAGE Collaboration, J. N. Abdurashitov  
 {\it et al.},  {\it Phys. Rev.} {\bf C 60}, 055801 (1999); V. N. Gavrin, 
 {\it Nucl. Phys.} {\bf 91} ({\it Proc. Suppl.}), 36 (2001); 
 Super-Kamiokande Collaboration, Y. Fukuda  {\it et al.}, 
 {\it Phys. Rev. Lett.} {\bf 81}, 1158 (1999), (E) 
 {\it ibid.}{\bf 81}, 4279 (1998); {\it ibid.}{\bf 82}, 1810 (1999);
 {\it ibid.}{\bf 82}, 2430 (1999); Y. Suzuki,   
 {\it Nucl. Phys.} {\bf 91} ({\it Proc. Suppl.}), 29 (2001); 
 SNO Collaboration, Q. R. Ahmad {\em et al.}, 
 {\it Phys. Rev. Lett.} {\bf 87}, 071301 (2001).

\bibitem{SNO}  SNO Collaboration, Q. R. Ahmad {\em et al.}, 
 {\it Phys. Rev. Lett.} {\bf 89}, 011301 (2002); {\it ibid.}{\bf 89}, 011302 (2002).  

\bibitem{Alicki} R. Alicki and K. Lendi, {\it Quantum Dynamical Semigroups 
and Applications}, Lect. Notes Phys. {\bf 286}, Springer-Verlag, Berlin 
 (1987).

\bibitem{Lindblad} G. Lindblad, {\it Commun. Math. Phys.} {\bf 48}, 119 
(1976).

\bibitem{Sudarshan} V. Gorini, A. Kossakowski and E. C. G. Sudarshan, 
{\it J. Math. Phys.} {\bf 17}, 821 (1976).

\bibitem{Gell-Mann} M. Gell-Mann,  {\it Phys. Rev.} {\bf 125}, 1067 (1962).

\bibitem{Benatti4} F. Benatti and H. Narnhofer, {\it Lett. Math. Phys.} 
 {\bf 15}, 325 (1988).

\bibitem{Gorini} V. Gorini, A. Frigerio, M. Verri, A. Kossakowski and 
E. C. G. Sudarshan, {\it Rep. Math. Phys.} {\bf 13}, 149 (1978).

\bibitem{Frigerio} A. Frigerio and V. Gorini, {\it J. Math. Phys.} 
 {\bf 17}, 2123 (1976).

\bibitem{Kossakowski} A. Kossakowski, {\it Rep. Math. Phys.} {\bf 3}, 
 247 (1972).

\bibitem{Ingarden} R. S. Ingarden and A. Kossakowski, {\it Ann. Phys.} 
 {\bf 89}, 451 (1975).

\bibitem{Davies2} E. B. Davies, {\it Commun. Math. Phys.} {\bf 39}, 
  91 (1974).

\bibitem{Martin} C. Favre and P. A. Martin, {\it Helv. Phys. Acta} 
 {\bf 41}, 333 (1968).

\bibitem{Spohn} R. D\"umcke and H. Spohn, {\it Z. Phys. B} {\bf 22}, 
 419 (1979).

\bibitem{chooz} CHOOZ Collaboration, M. Apollonio {\it et al.}, {\it Phys. Lett.} {\bf B 466}, 415 (1999).

\bibitem{k2k} K2K Collaboration, J. E. Hill {\it et al.}, hep-ex/0110034. 

\bibitem{shiozawa} Super-Kamiokande Collaboration, M. Shiozawa, talk at Neutrino 2002, http://neutrino2002.ph.tum.de/.  

\bibitem{kajita} T. Kajita and Y. Totsuka, {\it Rev. Mod. Phys.} {\bf 73}, 85 (2001).

\bibitem{smy} Super-Kamiokande Collaboration, S. Fukuda  {\it et al.}, {\it Phys. Lett.} {\bf B 539}, 179 (2002).

\bibitem{prem} A. Dziewonski and D. Anderson, {\it Phys. Earth Planet. Inter.} {\bf 25}, 297 (1981). 
\end{thebibliography}
\end{document}